\documentclass[aps,pra,twocolumn,superscriptaddress,nofootinbib,longbibliography]{revtex4-2}

\usepackage{amsmath,amssymb,bm}
\usepackage{graphicx}
\usepackage[colorlinks=true,citecolor=blue,linkcolor=blue,urlcolor=blue]{hyperref}

\begin{document}

\title{Detector-Conditioned Source-Space Nulls and Null-Mask Loss in a Programmable Two-Slit Interferometer}

\author{Jianming Wen}
\email{jwen7@binghamton.edu}
\affiliation{Department of Electrical and Computer Engineering, Binghamton University, Binghamton, New York 13902, USA}

\date{\today}

\begin{abstract}
Afshar's double-slit experiment probes wave--particle complementarity by placing a wire grid at the dark fringes of a downstream interference pattern while retaining an imaging basis that appears to preserve which-path information. Here we propose and analyze a time-reversed Young--Afshar configuration in which the corresponding null test is transferred from the downstream field plane to the source-label plane of a time-reversed Young interferometer. In this reciprocal geometry, a point-addressable source illuminates a double slit, while the detector remains fixed. The observed fringe is therefore not a single-shot spatial intensity pattern, but a detector-conditioned response reconstructed by scanning the source coordinate. Consequently, a null in this pattern is not a node of a freely propagating field; it is a source label for which the coherent two-slit transfer amplitude to the selected detector vanishes. A mask placed at such source-plane labels is invisible to that detector when both slits are open, yet becomes visible when either slit is opened alone. We develop the scalar Fresnel model, derive the source-space null condition, introduce a detector-conditioned null-mask loss, and examine how this loss evolves under a tunable which-path marker. The source-space visibility and path distinguishability satisfy the standard duality relation, so no violation of complementarity is implied. The essential new feature is instead a reciprocal, detector-conditioned form of complementarity: Afshar's field-space transparency is replaced by response-function transparency in a reconstructed source basis.
\end{abstract}

\maketitle

\section{Introduction}
Young's double-slit experiment remains the canonical demonstration of interference and wave--particle complementarity~\cite{Young,Taylor,Wooters,Bartell,Mittelstaedt,Greenberger,Scully,Storey,Jaeger,Chapman,Englert1996,Zeilinger,Kim,Walborn,Schouten,Mir,Kocsis,Menzel,Svensson,Jennewein,Aharonov,Xiao,Ketterle}. In the standard geometry, a fixed source illuminates two slits and produces an interference pattern in a downstream observation plane. Afshar's modification~\cite{Afshar2005,Afshar2006,Afshar2007} placed a thin wire grid at the dark fringes of this pattern while using a lens to image the two pinholes onto separated output spots. With both pinholes open, the grid produces little loss because it lies at interference minima, whereas with either pinhole closed, the same grid produces appreciable loss. Afshar interpreted this behavior as simultaneous evidence of wave interference and path information. Subsequent analyses have clarified, however, that no violation of complementarity occurs once visibility, distinguishability, and postselected subensembles are defined consistently \cite{Steuernagel2007,Jacques2008,Kastner,Flores}.

Here we ask a reciprocal question: what becomes of the Afshar logic in a time-reversed Young (TRY) geometry~\cite{Wen2025,WenHybrid,Wen2026a,Wen2026b}, where the detector is fixed and the source coordinate is scanned or programmed? In a TRY interferometer, the measured fringe is not a single-shot intensity distribution in an output plane. Instead, it is a detector-conditioned response reconstructed over the source coordinate. This distinction changes the meaning of a dark fringe. A conventional Young dark fringe is a spatial node of a propagated field, whereas a TRY dark fringe is an input source label whose coherent two-path transfer amplitude to a selected detector vanishes.

This distinction is more than a geometrical rearrangement. In the conventional Afshar experiment, the wire grid samples an already-formed downstream interference field. In the TRY-Afshar version proposed here, the mask is placed in the source plane before the double slit. It does not sit in a dark optical field; instead, it removes selected input preparations. The mask is invisible to the chosen detector only because, with both slits open, those source labels make zero coherent contribution to that detector. If either slit is closed, the destructive cancellation is removed, the same labels become bright in the detector-conditioned response, and the mask produces a finite loss.

We call this effect detector-conditioned source-space null-mask invisibility. It may be viewed as a source-space analog of Afshar transparency, but the analogy must be drawn carefully. The conventional Afshar grid probes field-space intensity nodes, whereas the TRY mask probes zeros of an input--output transfer function. The central aim of this paper is to formulate this distinction quantitatively and to show how it remains fully consistent with complementarity.

\begin{figure}[htbp]
\includegraphics[width=\columnwidth]{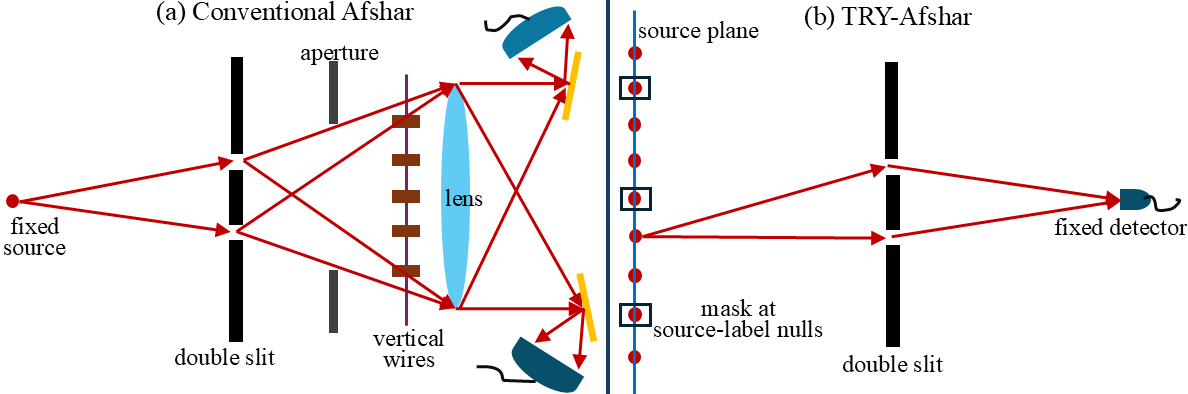}
\caption{Conceptual distinction between conventional Afshar interferometry and the proposed TRY-Afshar geometry. (a) In the conventional Afshar experiment, a wire grid is placed at the dark fringes of a downstream two-slit interference field before the slit images are formed by a lens. The grid probes true field-space nulls. (b) In the TRY-Afshar geometry, the detector is fixed and the source coordinate is scanned or programmed. A source-plane mask removes labels that reconstruct as nulls of the selected detector response. The mask therefore removes input preparations rather than sampling a downstream dark field.}
\label{fig:fig1}
\end{figure}

Figure~\ref{fig:fig1} summarizes the conceptual difference. In the conventional Afshar arrangement, the wire grid is placed after the double slit and samples dark fringes of a real propagated field. In the TRY-Afshar arrangement, the detector is fixed and the source coordinate is scanned or programmed. The mask is placed before the double slit and removes source labels that reconstruct as nulls only after coherent propagation to the selected detector. Thus the two experiments probe different objects: a field-space intensity node in the former case and a detector-conditioned transfer-function zero in the latter.

The paper is organized as follows. Section~\ref{sec:model} derives the two-slit input--output response in the Fresnel approximation. Section~\ref{sec:where} identifies where interference is generated in the TRY geometry. Section~\ref{sec:mask} introduces the source-plane null mask and derives the detector loss for two-slit and one-slit configurations. Section~\ref{sec:comp} adds a tunable which-path marker and shows that the TRY null fills in according to the standard visibility--distinguishability tradeoff. Section~\ref{sec:comparison} compares the proposed experiment with conventional Afshar interferometry and clarifies what is, and is not, new.

\section{Input--output model of the TRY interferometer}
\label{sec:model}
We use a scalar paraxial model to isolate the essential complementarity physics. A point-addressable source lies in the plane $z=-z_s$, with transverse coordinate $y$. Two narrow slits are located at $z=0$ with positions
\begin{equation*}
x_1=-\frac{a}{2},\quad x_2=+\frac{a}{2},
\end{equation*}
where $a$ is the slit separation. A detector is fixed in the plane $z=z_d$ at transverse coordinate $X$. The wavelength is $\lambda$ and the wavenumber is $k=2\pi/\lambda$.

For a source point $y$, the amplitude reaching the detector through slit $j$ is, apart from a common propagation factor,
\begin{eqnarray}
 A_j(y,X)=\alpha_j\exp\!\left[ik\frac{(x_j-y)^2}{2z_s}+ik\frac{(X-x_j)^2}{2z_d}+i\phi_j\right]\!.\label{eq:Aj}
\end{eqnarray}
Here $\alpha_j$ includes the slit transmission and slowly varying diffraction factors, while $\phi_j$ is an additional controllable path phase. Finite slit widths only multiply this expression~(\ref{eq:Aj}) by diffraction envelopes. They do not change the source-space null condition as long as the two paths can be balanced at the selected detector.

With both slits open and no which-path marking, the detector-conditioned response is
\begin{equation}
R_{12}(y,X)=\left|A_1(y,X)+A_2(y,X)\right|^2.\label{eq:R12def}
\end{equation}
Writing $A_j=\sqrt{I_j}\exp(i\theta_j)$, this becomes
\begin{equation}
R_{12}(y,X)=I_1+I_2+2\sqrt{I_1I_2}\cos\Phi(y,X),\label{eq:R12general}
\end{equation}
where $\Phi(y,X)=\theta_2(y,X)-\theta_1(y,X)$. Using the two slit positions, the paraxial phase difference is
\begin{align}
\Phi(y,X)=\Delta\phi-ka\left(\frac{y}{z_s}+\frac{X}{z_d}\right),\label{eq:phase}
\end{align}
with $\Delta\phi=\phi_2-\phi_1$. In the balanced case $I_1=I_2=I_0$, Eq.~(\ref{eq:R12general}) reduces to
\begin{eqnarray}
R_{12}(y,X)=2I_0\left[1+\cos\Phi(y,X)\right]=4I_0\cos^2\frac{\Phi(y,X)}{2}.
\label{eq:balancedR}
\end{eqnarray}

For a fixed detector position $X$, the dark source labels $y_m$ satisfy
\begin{equation}
\Phi(y_m,X)=(2m+1)\pi,\label{eq:nullphase}
\end{equation}
which gives
\begin{equation}
y_m=-\frac{z_s}{z_d}X+\frac{z_s}{ka}\left[\Delta\phi-(2m+1)\pi\right].\label{eq:ym}
\end{equation}
The spacing between adjacent source-space nulls is therefore
\begin{equation}
\Delta y=\frac{2\pi z_s}{ka}=\frac{\lambda z_s}{a}.\label{eq:dy}
\end{equation}
This is the source-space counterpart of the usual Young-fringe condition, but its physical meaning is different. It does not identify a dark point on an output screen. It identifies an input source coordinate whose coherent two-path transfer amplitude into a selected fixed detector mode vanishes.

\begin{figure}[htbp]
\includegraphics[width=0.96\columnwidth]{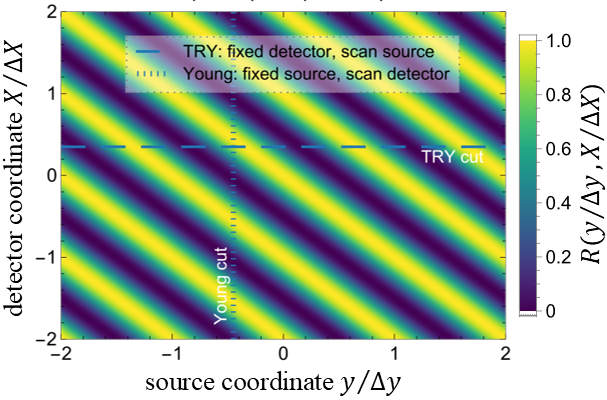}
\caption{Input--output response map of the normalized two-slit transfer function $R_{12}(y/\Delta y,X/\Delta X)=[1+\cos2\pi(y/\Delta y+X/\Delta X)]/2$. As an example, the vertical dotted line, $y/\Delta y=-0.45$, represents the conventional Young cut: the source is fixed and the detector coordinate is scanned. The horizontal dashed line, $X/\Delta X=0.35$, represents the TRY cut: the detector is fixed and the source coordinate is scanned. Although both cuts come from the same two-coordinate interference kernel, their nulls have different meanings. A Young null is a field-space node in the output plane, whereas a TRY null is a source label whose transfer amplitude into the selected detector mode vanishes.}
\label{fig:fig2}
\end{figure}

The joint source-detector response is illustrated in Fig.~\ref{fig:fig2} using the normalized two-slit kernel~\eqref{eq:balancedR}. A conventional Young experiment fixes the source coordinate and scans the detector coordinate, corresponding to a vertical cut through this two-coordinate response map. A TRY experiment fixes the detector coordinate and scans the source coordinate, corresponding to a horizontal cut. Thus the two arrangements sample the same propagation kernel, but they assign different operational meanings to its nulls. Specifically, a Young null is an output-plane field node; a TRY null is a detector-conditioned source-label zero.

\section{Where is the interference generated?}
\label{sec:where}
The TRY geometry raises a subtle question that does not arise as sharply in the conventional Young experiment: if the reconstructed fringe is indexed by source position, is the interference generated before or after the double slit?

The answer is that the TRY fringe is not localized exclusively on either side of the slits. The phase difference~\eqref{eq:phase} can be written separated into an input-side and an output-side contribution:
\begin{equation}
\Phi(y,X)=\Phi_{\rm in}(y)+\Phi_{\rm out}(X)+\Delta\phi,\label{eq:phaseparts}
\end{equation}
where
\begin{equation*}
\Phi_{\rm in}(y)=-ka\frac{y}{z_s},\quad\Phi_{\rm out}(X)=-ka\frac{X}{z_d}.
\end{equation*}
Thus, scanning the source coordinate changes the relative phase with which the two slits are illuminated. In this sense, the TRY fringe is controlled from the input side. However, the actual destructive cancellation described by Eq.~(\ref{eq:ym}) occurs only after the two slit alternatives propagate to, and are projected onto, the same fixed detector mode $X$.

Therefore, the TRY interference should not be ascribed as occurring solely before or solely after the double slit. Rather, it is a property of the full input--output transfer kernel
\begin{equation}
 H_X(y)=A_1(y,X)+A_2(y,X).\label{eq:kernel}
\end{equation}
The reconstructed detector-conditioned response~\eqref{eq:R12def} is
\begin{equation}
 R_{12}(y,X)=|H_X(y)|^2.
\end{equation}
A dark source label $y_m$ is therefore a zero of this transfer kernel $H_X(y)$ for the chosen detector mode $X$:
\begin{equation}
H_X(y_m)=0.\label{eq:kernelzero}
\end{equation}

This is the key distinction from conventional Young interference. In the usual experiment, the source coordinate $y$ is fixed and the detector coordinate $X$ is scanned or recorded; the fringe is a spatial intensity distribution in a downstream plane. In the TRY experiment, the detector coordinate $X$ is fixed and the source coordinate $y$ is scanned or programmed; the fringe is a reconstructed response over input labels. The two cases correspond to different slices of the same two-coordinate propagation kernel~(\ref{eq:kernel}) (as schematically illustrated Fig.~\ref{fig:fig2}), but they are not operationally equivalent.

For a single source label $y_m$, the condition $R_{12}(y_m,X)=0$ means only that this preparation produces destructive interference at the selected detector. It does not imply that the source plane is dark, nor that the entire field behind the slits vanishes. The full TRY fringe emerges only after repeated trials over many source labels. For this reason, a source-plane mask in the TRY-Afshar geometry cannot be interpreted as a wire placed at a physical dark fringe. It is instead a preparation filter: it removes source labels whose coherent transfer amplitude to one chosen detector channel is zero.

\section{Source-plane null mask}
\label{sec:mask}

We now introduce the TRY analog of the Afshar grid. Let $S$ denote the set of source labels that coincide with reconstructed two-slit nulls for the selected detector position $X$:
\begin{equation}
S=\{y_m:\; H_X(y_m)=0\}.
\end{equation}
A source-plane mask removes these labels. For a discrete source array, this mask can be written as
\begin{equation}
M_S(y_n)=\begin{cases}
0, & y_n\in S,\\
1, & y_n\notin S.
\end{cases}\label{eq:maskdiscrete}
\end{equation}
For a continuous source distribution, the ideal discrete mask~(\ref{eq:maskdiscrete}) may be replaced by a finite-width mask,
\begin{equation}
M_S(y)=1-\eta\sum_mg_w(y-y_m),\label{eq:maskcontinuous}
\end{equation}
where $0\leq\eta\leq 1$ is the opacity and $g_w$ is a localized window of width $w$ centered at a null label $y_m$.

The observable must be defined at the ensemble level. For a single ideal null label $y_m$, the two-slit detector count is zero both with and without the mask, so a local transmission ratio of the two is not meaningful. The physically meaningful quantity is instead the total count loss over a programmed source ensemble.

Let $p_n$ be the probability, or equivalently the dwell-time weight, assigned to source label $y_n$. Without the mask, the selected-detector count in the two-slit configuration is
\begin{equation}
N_{12}^{(0)}=\sum_np_nR_{12}(y_n,X).\label{eq:N120}
\end{equation}
With the source mask present,
\begin{equation}
N_{12}^{(M)}=\sum_np_nM_S(y_n)R_{12}(y_n,X).\label{eq:N12M}
\end{equation}
The mask-induced two-slit loss is therefore
\begin{equation}
\Delta N_{12}=N_{12}^{(0)}-N_{12}^{(M)}=\sum_{y_n\in S}p_n R_{12}(y_n,X).
\label{eq:loss12}
\end{equation}
If the selected source labels are exact reconstructed two-slit nulls, then
\begin{equation}
R_{12}(y_n,X)=0,\quad y_n\in S,
\end{equation}
and hence
\begin{equation}
\Delta N_{12}=0.\label{eq:loss12zero}
\end{equation}
Thus, in the ideal limit, the source-plane mask is invisible to the selected detector in the coherent two-slit configuration. This invisibility is not global; it is conditioned on both the two-slit interference and the chosen detector channel.

Now close slit 2. The detector response becomes
\begin{equation}
R_1(y,X)=|A_1(y,X)|^2.
\end{equation}
The corresponding mask-induced loss is
\begin{equation}
\Delta N_1=\sum_{y_n\in S}p_n|A_1(y_n,X)|^2.\label{eq:loss1}
\end{equation}
There is no interference cancellation in this one-slit configuration, so this loss generally does not disappear. In the balanced and slowly varying limit, 
\[
|A_1(y_n,X)|^2\approx I_0, 
\]
and therefore
\begin{equation}
\Delta N_1\approx I_0\sum_{y_n\in S}p_n.\label{eq:loss1approx}
\end{equation}
Similarly, with only slit 2 open,
\begin{equation}
\Delta N_2=\sum_{y_n\in S}p_n|A_2(y_n,X)|^2\approx I_0\sum_{y_n\in S}p_n.
\label{eq:loss2}
\end{equation}
The same source labels are therefore dark for the coherent two-slit transfer into the selected detector, but bright for either single-slit transfer. This is the essential TRY--Afshar signature.

\begin{figure}[htbp]
\includegraphics[width=0.85\columnwidth]{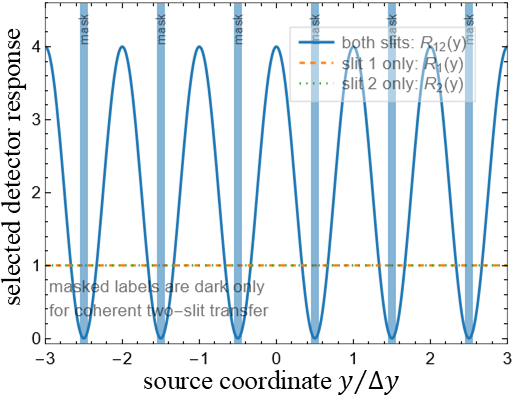}
\caption{Source-space nulls and detector-conditioned null-mask invisibility.
The coherent two-slit response is plotted as
$R_{12}(y)=2[1+\cos(2\pi y/\Delta y)]$, while the one-slit responses are taken as $R_1(y)=R_2(y)=1$ over the displayed source window. The shaded regions mark source-plane mask positions centered at the reconstructed two-slit null labels $y/\Delta y=n+1/2$. With both slits open, these labels have zero transfer amplitude to the selected detector and therefore produce negligible mask-induced loss. With either slit alone, the same labels are bright, so their removal produces a finite detector loss.}
\label{fig:fig3}
\end{figure}

Figure~\ref{fig:fig3} illustrates this detector-conditioned null-mask behavior. The coherent two-slit response contains reconstructed source-space nulls at the labels selected by the mask. Removing those labels produces no loss in the ideal two-slit response because they already have zero transfer amplitude to the chosen detector. With either slit alone, however, the cancellation is absent. The same labels then contribute finite one-path amplitudes, and the mask produces a measurable loss. The mask is consequently invisible only to the coherent two-slit channel selected by the detector; it is not a universally transparent object.

A compact experimental contrast can be defined as
\begin{equation}
\mathcal{C}_{\rm N}=\frac{\Delta N_1+\Delta N_2}{2\Delta N_{12}+N_{\rm bg}},
\label{eq:Cnull}
\end{equation}
where $N_{\rm bg}$ accounts for background counts, imperfect extinction, and finite null depth. Ideally, $\Delta N_{12}\rightarrow0$ while $\Delta N_1$ and $\Delta N_2$ remain finite, giving $\mathcal{C}_{\rm N}\gg1$.

This is the corrected TRY-Afshar signature. The mask is not placed in a dark optical field. Instead, it removes input preparations whose coherent two-path transfer amplitude into the selected detector channel is zero.

\subsection{Shifted-mask test}
A direct way to verify the source-space nature of the effect is to shift the mask away from the reconstructed nulls. Let
\begin{equation}
M_S(y;\delta)=1-\eta\sum_m g_w(y-y_m-\delta).
\end{equation}
For a narrow mask and slowly varying source weights, the two-slit loss samples the response near $y_m+\delta$:
\begin{equation}
\Delta N_{12}(\delta)\propto \sum_m p_m R_{12}(y_m+\delta,X).
\end{equation}
Using the source-space phase relation~(\ref{eq:phase}), we have
\begin{equation}
\Phi(y_m+\delta,X)=(2m+1)\pi-\kappa\delta,\quad\kappa=\frac{ka}{z_s}.
\end{equation}
For balanced paths,
\begin{align}
R_{12}(y_m+\delta,X)
=2I_0\left[1-\cos(\kappa\delta)\right].\label{eq:shiftloss}
\end{align}
Thus, near a null,
\begin{equation}
R_{12}(y_m+\delta,X)\approx I_0\kappa^2\delta^2,\quad|\kappa\delta|\ll1.
\end{equation}
The two-slit loss therefore rises quadratically as the mask is displaced away from the reconstructed null. For a finite-width source mask, this local quadratic response gives a two-slit loss that scales as $\kappa^2w^2$ relative to the one-slit loss, as derived in Appendix~\ref{app:finite_mask}. By contrast, in either single-slit configuration, the shifted-mask loss remains approximately independent of $\delta$ as long as the single-slit envelope is locally flat. This shifted-mask test cleanly distinguishes a coherent detector-conditioned source-space null from an ordinary modulation of source intensity.

\section{Complementarity in the source-space null basis}
\label{sec:comp}

We now introduce a tunable which-path marker. Suppose that passage through slit $j$ correlates the optical field with a normalized marker state $|m_j\rangle$. If this marker degree of freedom is not measured at the detector, the detected response is obtained from the coherent field--marker superposition
\begin{equation*}
A_1(y,X)|m_1\rangle+A_2(y,X)|m_2\rangle.
\end{equation*}
The resulting two-slit response is
\begin{equation}
R_{12}^{(\mu)}(y,X)=I_1+I_2+2\sqrt{I_1I_2}\,\mathrm{Re}\left[\mu e^{i\Phi(y,X)}\right],\label{eq:Rmu}
\end{equation}
where
\begin{equation}
\mu=\langle m_1|m_2\rangle
\end{equation}
is the marker overlap. The magnitude $|\mu|$ measures the remaining path coherence. For example, using polarization markers,
\begin{equation}
|m_1\rangle=|H\rangle,\quad|m_2\rangle=\cos\chi\,|H\rangle+\sin\chi\,|V\rangle,
\end{equation}
gives
\begin{equation}
\mu=\cos\chi.
\end{equation}

For balanced paths ($I_1=I_2=I_0$) and taking $\mu$ real and nonnegative for clarity, the response~(\ref{eq:Rmu}) becomes
\begin{equation}
R_{12}^{(\mu)}(y,X)=2I_0\left[1+\mu\cos\Phi(y,X)\right].\label{eq:Rbalancedmu}
\end{equation}
The source-space fringe visibility is therefore
\begin{equation}
V_s=\frac{R_{\max}-R_{\min}}{R_{\max}+R_{\min}}=|\mu|.\label{eq:Vs}
\end{equation}
The path distinguishability associated with the two marker states is
\begin{equation}
D=\sqrt{1-|\mu|^2}.\label{eq:Dbalanced}
\end{equation}
Hence,
\begin{equation}
V_s^2+D^2=1.\label{eq:dualitybalanced}
\end{equation}
This is the standard complementarity relation, now expressed not in a detector-plane fringe basis, but in the reconstructed source-space response basis.

\begin{figure}[htbp]
\includegraphics[width=0.85\columnwidth]{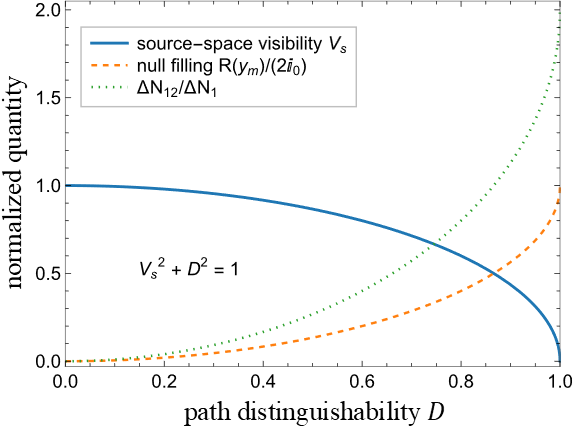}
\caption{Complementarity through source-space null filling. For balanced paths with marker overlap $\mu=\sqrt{1-D^2}$, the source-space visibility is $V_s=\sqrt{1-D^2}$, so $V_s^2+D^2=1$. A source label $y_m$ that is a perfect null for indistinguishable paths fills in according to
$R(y_m)/(2I_0)=1-\sqrt{1-D^2}$. The corresponding relative null-mask loss follows $\Delta N_{12}/\Delta N_1=2[1-\sqrt{1-D^2}]$. Increasing path distinguishability therefore makes the source-plane mask visible to the selected detector by destroying the coherent source-space transfer zero.}
\label{fig:fig4}
\end{figure}

Figure~\ref{fig:fig4} illustrates this tradeoff. As the path distinguishability $D$ increases, the source-space visibility $V_s$ decreases according to the usual duality relation. At the same time, a source label that was a perfect two-slit null for indistinguishable paths becomes bright. Thus the appearance of mask-induced loss in the two-slit configuration is not a violation of complementarity. It is a direct signature of complementarity: which-path information destroys the coherent transfer zero and fills in the source-space null.

Now consider a source label $y_m$ that is a perfect null when the marker states are identical, $\mu=1$. At this label, $\Phi(y_m,X)=(2m+1)\pi$, so Eq.~(\ref{eq:Rbalancedmu}) gives
\begin{equation}
R_{12}^{(\mu)}(y_m,X)=2I_0(1-\mu).\label{eq:nullfill}
\end{equation}
The former source-space null fills in continuously as which-path information is introduced. Since $\mu=\sqrt{1-D^2}$, this can be written as
\begin{equation}
R_{12}^{(D)}(y_m,X)=2I_0\left(1-\sqrt{1-D^2}\right).\label{eq:nullfillD}
\end{equation}
For weak marking, $D\ll1$,
\begin{equation}
R_{12}^{(D)}(y_m,X)\approx I_0D^2.
\end{equation}
For complete marking, $D=1$ and $\mu=0$, the former null reaches the incoherent two-slit value,
\begin{equation}
R_{12}^{(D=1)}(y_m,X)=2I_0.
\end{equation}

The source-mask loss follows the same null-filling law. For a mask placed at labels that are exact nulls when $\mu=1$,
\begin{equation}
\Delta N_{12}^{(\mu)}=\sum_{y_n\in S}p_nR_{12}^{(\mu)}(y_n,X).
\end{equation}
Using the balanced-path result,
\begin{equation}
\Delta N_{12}^{(\mu)}=2I_0(1-\mu)\sum_{y_n\in S}p_n.\label{eq:lossmu}
\end{equation}
The corresponding one-slit loss remains approximately
\begin{equation}
\Delta N_1\approx I_0\sum_{y_n\in S}p_n.
\end{equation}
Thus the normalized two-slit loss relative to the single-slit loss is
\begin{equation}
\frac{\Delta N_{12}^{(\mu)}}{\Delta N_1}\approx 2(1-\mu)=2\left(1-\sqrt{1-D^2}\right).\label{eq:lossratioD}
\end{equation}
This relation is experimentally useful. The source-plane mask is invisible to the selected detector only when the two paths remain mutually coherent and indistinguishable. As which-path information increases, the mask becomes visible because the destructive source-space transfer zero is progressively erased.

For completeness, consider unequal path intensities $I_1$ and $I_2$. The source-space visibility becomes
\begin{equation}
V_s=\frac{2\sqrt{I_1I_2}|\mu|}{I_1+I_2}.\label{eq:Vunbalanced}
\end{equation}
For pure marker states, the corresponding distinguishability is
\begin{equation}
D=\sqrt{1-\frac{4I_1I_2|\mu|^2}{(I_1+I_2)^2}},\label{eq:Dunbalanced}
\end{equation}
so that again
\begin{equation}
V_s^2+D^2=1.\label{eq:dualitygeneral}
\end{equation}
A perfect source-space null requires both phase opposition and amplitude balance. If $I_1\neq I_2$, the minimum response is
\begin{equation}
R_{\min}=I_1+I_2-2|\mu|\sqrt{I_1I_2}.
\end{equation}
Thus partial coherence, partial imbalance, and finite marker distinguishability, and background counts all appear as finite null depth. The TRY null-mask loss is therefore not only a complementarity test, but also a sensitive diagnostic of coherence, alignment, and path balance in the source-space response basis.

\section{Comparison with conventional Afshar interferometry}
\label{sec:comparison}

The proposed experiment is inspired by Afshar's wire-grid logic, but its operational meaning is different.

In conventional Young--Afshar interferometry, the two slits generate a real spatial interference field downstream. A wire placed at a dark fringe samples a location where the propagated intensity is nearly zero:
\begin{equation*}
I(x_{\rm wire})\approx0.
\end{equation*}
Its low loss is therefore a field-space statement: the wire lies at a physical intensity minimum of the propagated optical field.

In the TRY-Afshar geometry the reconstructed fringe is not an output-plane intensity pattern. It is the detector-conditioned response $R_{12}(y,X)$, obtained by scanning or programming the source label $y$ while keeping the detector fixed at $X$. A TRY null is therefore not a dark point in the source plane. It is a zero of the input-output transfer function:
\begin{equation*}
H_X(y_m)=A_1(y_m,X)+A_2(y_m,X)=0.
\end{equation*}
The source-plane mask does not block a dark field. It removes an input preparation that would have produced no count in the selected detector channel.

This distinction also makes the transparency conditional. The mask is invisible only for the chosen detector and only when both slits are open and mutually coherent. The same removed source labels may contribute to other detector positions, other output modes, or to the same detector when either slit is used alone. Thus the TRY--Afshar signature is
\begin{equation*}
\Delta N_{12}\approx0,\quad\Delta N_1,\Delta N_2>0.
\end{equation*}

The complementarity question is therefore shifted. In conventional Afshar discussions, the issue is whether a high-visibility downstream interference pattern can be inferred while retaining path information in an image basis. In the TRY version, the wave-like witness is the reconstructed source-space response, while the particle-like information is introduced through path marking or output-mode distinguishability. The relevant tradeoff is therefore between source-space visibility and path distinguishability, and it obeys the standard duality relation
\[
V_s^2+D^2\le1,
\]
with equality for ideal pure marker states.

Thus the proposed experiment should not be presented as a violation of complementarity. Its novelty is more precise and, in some ways, cleaner: Afshar's field-space transparency is transformed into detector-conditioned response-function transparency. The resulting null-mask loss then directly shows how a source-space interference zero is created by coherent path indistinguishability and destroyed by which-path marking.

\section{Experimental considerations}

A proof-of-principle experiment can be implemented with weak coherent light. The source plane may be realized using a scanned single-mode fiber, a digital micromirror device, a spatial light modulator, or an addressable emitter array. The double slit is placed a distance $z_s$ from the source plane, and a single-pixel detector is fixed at a chosen output coordinate $X$. 

The basic procedure is straightforward:
\begin{enumerate}
\item Scan the source coordinate $y$ and record the two-slit response $R_{12}(y,X)$.
\item Identify the reconstructed dark source labels $S=\{y_m\}$ for the selected detector.
\item Apply a physical or programmable source-plane mask that suppresses the labels in $S$.
\item Measure the detector loss with both slits open and then with each slit open individually.
\item Shift the mask by a controlled displacement $\delta$ and verify that the two-slit loss rises away from the null according to the predicted modulation (Eq.~(\ref{eq:shiftloss})).
\item Introduce a tunable which-path marker, for example by rotating the polarization in one slit path, and measure the corresponding filling of the source-space null (Eq.~(\ref{eq:nullfill})).
\end{enumerate}

A single-photon version can be realized either by attenuating a coherent source or by using heralded photons. In a heralded implementation, the idler photon may label or gate the effective source coordinate, while the signal photon propagates through the double slit. Coincidence conditioning then reconstructs $R_{12}(y,X)$. This version would make the preparation-label nature of the null mask especially clear, because the relevant ``dark labels" are defined by detector-conditioned coincidences rather than by a local intensity minimum in the source plane. The same complementarity physics applies: path marking reduces the source-space visibility and fills the reconstructed null.

Several practical imperfections should be included in a quantitative fit. Finite slit width introduces slowly varying diffraction envelopes into $I_1(y,X)$ and $I_2(y,X)$. Finite mask width averages the response around each null; since the ideal two-slit response rises quadratically with displacement from a null, the integrated loss scales as the third power of the mask width in the small-window limit; see Appendix~\ref{app:finite_mask}. Detector dark counts, imperfect extinction, phase noise, and unequal path amplitudes all produce a finite residual background and limit the achievable null depth (Eq.~(\ref{eq:Cnull})). These effects reduce the contrast but do not change the essential signature: the mask produces negligible loss for the coherent two-slit transfer channel, yet finite loss when either slit is used alone.

\section{Discussion}
The central significance of the proposed experiment is not the revival of Afshar's original paradox, but a sharper operational distinction between \emph{field-space interference} and \emph{source-space response interference}.

In an ordinary Young experiment, a dark fringe has a direct spatial meaning: it is a point in the output plane where the propagated intensity is small. In a TRY experiment, a dark label has a conditional transfer meaning: it is an input coordinate whose two-path amplitude has zero projection onto a selected detector mode. The reconstructed fringe is therefore not located simply before or after the double slit. The phase is controlled by the source coordinate, but the cancellation is completed only after propagation to the fixed detector. The fringe belongs to the full input--output Green function.

This viewpoint gives a natural source-space formulation of complementarity. The source coordinate acts as a programmable preparation basis, while the fixed detector selects an output mode. A source label that is bright through either slit alone can become dark only when the two paths are coherently indistinguishable. The dark label is therefore not a passive field node; it is an interference zero created by coherent path superposition and selected by the detector.

The null-mask experiment makes this statement directly measurable. When the paths are indistinguishable, masking the reconstructed null labels produces negligible loss at the selected detector. When which-path information is introduced, the same labels become bright and the mask produces a finite loss. The loss is therefore not a loophole in complementarity. It is a sensitive witness of complementarity: path distinguishability destroys the source-space interference zero.

This also explains why the TRY-Afshar experiment is not merely a time-reversed copy of Afshar’s original arrangement. In the conventional Afshar geometry, the wire grid samples an already-existing downstream field null. In the TRY geometry, the source mask removes input labels whose darkness is defined only after coherent propagation through both slits and projection onto a chosen detector. The null is relational: it belongs to a particular source–detector pair, not to the source plane alone.

This relational character is the most distinctive conceptual feature. A source label $y_m$ may be dark for one detector coordinate $X$ but bright for another detector coordinate $X'$. Likewise, it may be dark when both slits are open, yet bright when either slit is used alone. Thus, a TRY null is not an intrinsic property of a source point. It is a property of the complete interferometric transfer function.

The complementarity message is therefore more precise than in the traditional Afshar debate. The wave-like information is not a directly photographed interference pattern; it is the reconstructed source-space response and, in particular, the existence of detector-conditioned transfer zeros. The particle-like information is the ability to distinguish the two paths, whether by closing one slit, imaging the slits, or introducing a path marker. As this distinguishability increases, the transfer zeros fill in and the null-mask loss grows. Complementarity is not challenged; rather, it is displayed in a different and more explicitly conditional basis.

\section{Conclusion}
We have proposed a time-reversed Young--Afshar interferometer in which Afshar's downstream wire grid is replaced by a source-plane mask placed at detector-conditioned null labels. The distinction is essential. The mask does not sample dark fringes of a propagated field. Instead, it removes input preparations whose coherent two-slit transfer amplitude to a selected fixed detector vanishes. With both slits open, these source labels make no contribution to that detector; with either slit alone, they become bright. The result is a strong contrast between the two-slit and one-slit null-mask losses.

Introducing a tunable which-path marker shows that this effect is fully consistent with complementarity. As the path distinguishability increases, the source-space visibility decreases and the reconstructed null fills in according to the standard duality relation. The mask therefore becomes visible precisely when which-path information destroys the coherent transfer zero.

Thus, the proposed experiment does not challenge complementarity. In contrast, it reformulates it in a detector-conditioned source-space basis. Wave-like behavior appears as a reconstructed transfer-function null, while partile-like behavior appears as path distinguishability. In this sense, the TRY--Afshar geometry converts Afshar's field-space transparency into response-function transparency and provides a concrete operational test of complementarity in time-reversed Young interferometry.

\begin{acknowledgments}
The author acknowledges helpful discussions with Drs. Sidong Lei and Yanhua Zhai. This work was partially supported by Binghamton University through startup funds, Watson College through an internal award (105383), and NSF ExpandQISE-2329027.
\end{acknowledgments}

\appendix

\section{Finite-width source mask near a null}
\label{app:finite_mask}

For a continuous source coordinate, consider a single source-plane mask window centered at a reconstructed null $y_m$. Let $u=y-y_m$, and take the mask to be a top-hat window of width $w$, unit opacity, and locally constant source weight $p_m=p(y_m)$. Near the null, the balanced two-slit response is
\begin{equation}
R_{12}(y_m+u,X)=2I_0\left[1-\cos(\kappa u)\right]\approx I_0\kappa^2 u^2,
\end{equation}
where $\kappa=ka/z_s$. The mask-induced two-slit loss from this window is therefore
\begin{equation}
\Delta N_{12}^{(w)}\approx p_m I_0\kappa^2\int_{-w/2}^{w/2}u^2du=p_m I_0\kappa^2\frac{w^3}{12}.
\end{equation}
With only one slit open, the response is locally flat, so the corresponding loss is
\begin{equation}
\Delta N^{(w)}_1\approx p_mI_0\int^{w/2}_{-w/2}du=p_mI_0w.
\end{equation}
Thus, the ratio scales as
\begin{equation}
\frac{\Delta N_{12}^{(w)}}{\Delta N_1^{(w)}}\approx\frac{\kappa^2w^2}{12}.
\end{equation}
The two-slit loss is therefore quadratically suppressed by the mask width, while the one-slit loss scales linearly with width. In the narrow-mask limit, the continuous result approaches the ideal discrete-label result: a mask placed exactly at source-space null labels produces no loss in the coherent two-slit channel (see Eq.~(\ref{eq:loss12zero})).

\begin{figure}[h]
\includegraphics[width=0.85\columnwidth]{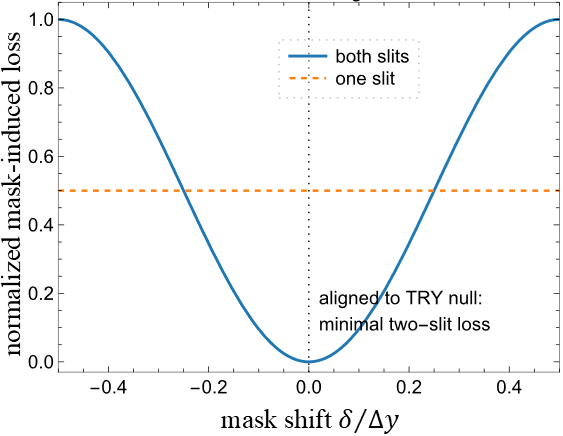}
\caption{Shifted-mask diagnostic of a source-space null. The two-slit loss is plotted as $R_{12}(\delta)=[1-\cos(2\pi\delta/\Delta y)]/2$, where $\delta$ is the displacement of a narrow mask from a reconstructed TRY null. The one-slit loss is shown as a constant reference, $R_1=0.5$, valid when the single-slit envelope varies slowly over the displacement range. The minimum at $\delta=0$ confirms that the mask is aligned with source labels whose coherent transfer amplitude to the selected detector vanishes.}
\label{fig:figS1}
\end{figure}

A useful control is to shift the mask away from the reconstructed null. As shown in Fig.~\ref{fig:figS1}, if the mask center is displaced by $\delta$, then for a narrow window the two-slit loss samples
\[
R_{12}(y_m+\delta,X)=2I_0[1-\cos(\kappa\delta)].
\]
Near the null,
\[
R_{12}(y_m+\delta,X)\approx I_0\kappa^2\delta^2.
\]
Thus the two-slit mask loss is minimized when the mask is aligned with the TRY null and increases quadratically as the mask is displaced. In contrast, the one-slit loss remains approximately constant over the same range, provided the one-slit diffraction envelope is locally flat. This shifted-mask test cleanly distinguishes coherent source-space nulling from ordinary attenuation of the source ensemble.

\end{document}